\documentclass[aps,prl,twocolumn,superscriptaddress,longbibliography]{revtex4-2}

\usepackage{gensymb}
\usepackage{amssymb}

\usepackage{xcolor}
\usepackage[normalem]{ulem}
\usepackage{graphicx}
\usepackage{dcolumn}
\usepackage{float}
\usepackage{bm}
\usepackage{amsmath}
\usepackage{comment}

\newcommand{\be}{\begin{eqnarray}}
\newcommand{\ee}{\end{eqnarray}}
\newcommand{\bfig}{\begin{figure}}
\newcommand{\efig}{\end{figure}}

\newcommand{\m}{\mathbf}

\begin{document}

\title{Frustration of harmonic and solitonic helimagnetism\\ on the body-centered tetragonal lattice of GdAlSi}

\author{Ryota Nakano}
\email{ryotagreg1023@g.ecc.u-tokyo.ac.jp}
\author{Rinsuke Yamada}
\affiliation{Department of Applied Physics and Quantum-Phase Electronics Center, The University of Tokyo, 113-8656, Tokyo, Japan}

\author{Oleg I. Utesov}
\email{utiosov@gmail.com}
\affiliation{Department of Physics, Korea Advanced Institute of Science and Technology, Daejeon 34141, Republic of Korea}

\author{Masaki Gen}
\affiliation{RIKEN Center for Emergent Matter Science (CEMS), Wako, Saitama 351-0198, Japan}
\affiliation{Institute for Solid State Physics, The University of Tokyo, Kashiwa 277-8581, Japan}

\author{Akiko Kikkawa}
\affiliation{RIKEN Center for Emergent Matter Science (CEMS), Wako, Saitama 351-0198, Japan}

\author{Hajime Sagayama}
\affiliation{Institute of Materials Structure Science, High Energy Accelerator Research Organization, Tsukuba, Ibaraki 305-0801, Japan}
\affiliation{Nagoya University Synchrotron Radiation Research Center (NUSR), Nagoya University, Nagoya 464-8603, Japan}

\author{Hironori Nakao}
\affiliation{Institute of Materials Structure Science, High Energy Accelerator Research Organization, Tsukuba, Ibaraki 305-0801, Japan}

\author{Masashi Tokunaga}
\affiliation{RIKEN Center for Emergent Matter Science (CEMS), Wako, Saitama 351-0198, Japan}
\affiliation{Institute for Solid State Physics, The University of Tokyo, Kashiwa 277-8581, Japan}

\author{Taka-hisa Arima}
\affiliation{RIKEN Center for Emergent Matter Science (CEMS), Wako, Saitama 351-0198, Japan}
\affiliation{Department of Advanced Materials Science, The University of Tokyo, Kashiwa 277-8561, Japan}

\author{Yoshinori Tokura}
\affiliation{Department of Applied Physics and Quantum-Phase Electronics Center, The University of Tokyo, 113-8656, Tokyo, Japan}
\affiliation{RIKEN Center for Emergent Matter Science (CEMS), Wako, Saitama 351-0198, Japan}
\affiliation{Tokyo College, The University of Tokyo, Tokyo 113-8656, Japan}

\author{Se Kwon Kim}
\affiliation{Department of Physics, Korea Advanced Institute of Science and Technology, Daejeon 34141, Republic of Korea}
\affiliation{Graduate School of Quantum Science and Technology, Korea Advanced Institute of Science and Technology, Daejeon 34141, Republic of Korea}

\author{Max Hirschberger}
\email{hirschberger@ap.t.u-tokyo.ac.jp}
\affiliation{Department of Applied Physics and Quantum-Phase Electronics Center, The University of Tokyo, 113-8656, Tokyo, Japan}
\affiliation{RIKEN Center for Emergent Matter Science (CEMS), Wako, Saitama 351-0198, Japan}

\date{\today}

\begin{abstract}
The triangular lattice antiferromagnet (TLAF) with nearest-neighbor exchange interaction is a model platform in the field of frustrated magnetism. Here, anharmonic (`up-up-down') and harmonic (`120 degree') magnetic states compete, because 
the fundamental helimagnetic wave and its higher harmonic are degenerate in energy.
We show that a body-centered tetragonal lattice (BCTL) can realize a similar frustration of harmonic and anharmonic helimagnetic states, and that 
the tetragonal magnetic Weyl semimetal GdAlSi realizes this scenario. In an applied magnetic field, resonant elastic X-ray scattering reveals a competition of harmonic cycloidal and solitonic double-$\m{Q}$ states,
well consistent with mean-field calculations. Our work provides a new paradigm for frustration physics in 
BCTL materials. 
\end{abstract} 

\maketitle
\textit{Introduction}~--~
Real-world materials with crystal lattices that harbor geometric frustration of magnetic interactions 
have attracted significant attention~\cite{Zhou_2017_QSL,moessner_2006_geometrical,balents_2010_SL,lacroix_2011_Frustration,ramirez1999zero,bramwell_2001_spinice,gingras_2014_spinice,zhao_2020_kagomespinice}.
In frustrated systems, competing interactions or lattice geometry prevent the simultaneous minimization of all local exchange energies, leading to degeneracy and enhanced spin fluctuations. 

As frustration provides a versatile mechanism for generating unconventional ground states~\cite{moessner_2006_geometrical,balents_2010_SL,lacroix_2011_Frustration}, extending its concepts beyond canonical lattice geometries is important. So far, studies of frustration-induced competing magnetic orders have been centered primarily on lattices with triangle plaquettes, such as the triangular, kagome, and pyrochlore lattices~\cite{mendels2016quantum,gingras2014quantum,tokura2025metal}. In the next paragraphs, 
we first revisit the essential frustration physics of the triangular lattice antiferromagnet (TLAF) before demonstrating analogous frustrated behavior on the body-centered tetragonal lattice (BCTL), which is the focus of this Letter.

\begin{figure*}[!htp]
  \centering
  \includegraphics[width=1.0\linewidth]{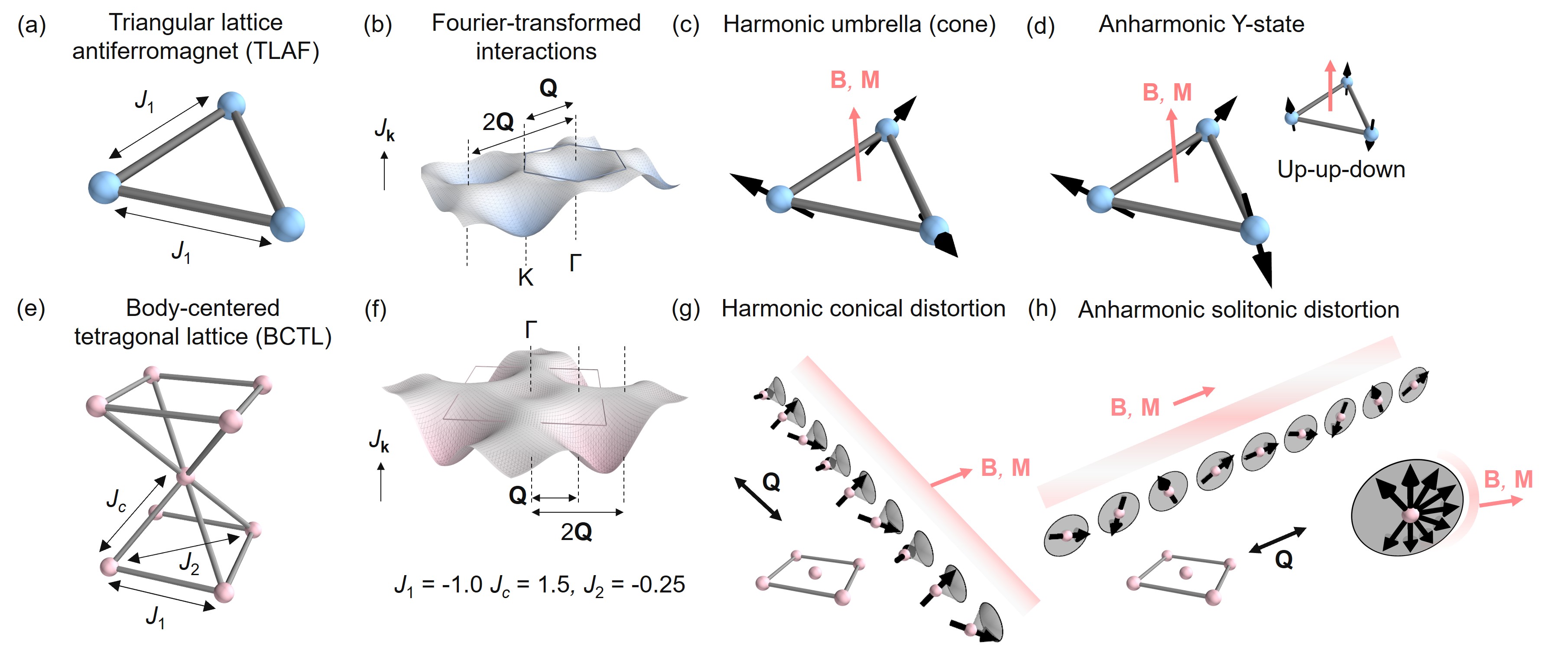} 
  \caption{(color online)
  Frustration and the resulting competing magnetic orders, for a triangular lattice antiferromagnet (TLAF) and a body-centered tetragonal lattice (BCTL).
(a) TLAF with geometric frustration induced by the nearest-neighbor antiferromagnetic interaction $J_1$.
(b) Fourier-transformed exchange interaction $J_{\m{k}}$ of the TLAF, exhibiting maxima at the equivalent K points in the hexagonal Brillouin zone (BZ, gray hexagon). This generically satisfies the condition $J_{\m{Q}} = J_{2\m{Q}}$. 
(c,d) Competing harmonic (umbrella, cone) and anharmonic (Y-state) magnetic orders on the TLAF under a magnetic field. Inset: up-up-down state, corresponding to the higher field limit of the Y-state.
(e) BCTL with competing nearest intralayer ($J_1$), nearest interlayer ($J_c$), and next-nearest-neighbor ($J_2$) interactions.
(f) Fourier-transformed exchange interactions for the BCTL, showing a maximum at $\m{Q} = (2/3,\,2/3,\,0)$ for $J_1 = -1.0$, $J_c = 1.5$, and $J_2 = -0.25$. The condition $J_{\m{Q}} = J_{2\m{Q}}$ is realized, similar to the TLAF.
(g,h) Competing conical (g) and solitonic (h) distortions of helimagnets under a magnetic field applied in the basal plane of the BCTL. The former (latter) acquires a net magnetization via uniform (non-uniform) spin canting. Insets (left), Projected effective 2D square lattice. Inset (right), anharmonic spin rotation associated with the solitonic distortion.
  }
  \label{Fig1}
\end{figure*}

On the TLAF, nearest-neighbor antiferromagnetic interactions cannot be simultaneously satisfied by a collinear configuration of magnetic moments [Fig.~\ref{Fig1}(a)]. As a consequence, frustration gives rise to a variety of competing ordered states, including noncollinear and noncoplanar spin textures, as well as superpositions of multiple magnetic waves: multi-$\m{Q}$ states.
The ground state is a three-sublattice structure in real space, corresponding to the ordering vector $\m{Q}=(1/3,\,1/3,\,0)$, which coincides with the $K$ point in reciprocal space. At $K$ and symmetry equivalent points, the Fourier-transformed exchange interaction $J_{\m{k}}$ is maximized, see Fig.~\ref{Fig1}(b). Importantly, the TLAF with nearest-neighbor interactions satisfies $J_{\m{Q}} = J_{2\m{Q}}$, which underlies its rich magnetic phase diagram.

Under an applied magnetic field, the degeneracy of the TLAF leads to multiple competing states, including the umbrella (cone) and Y-states shown in Fig.~\ref{Fig1}(c,d). The former (latter) corresponds to adding a uniform (nonuniform) magnetization component to the $120^\circ$ noncollinear ground state. With increasing field, the Y-state continuously evolves into an up-up-down configuration, giving rise to the characteristic $1/3$-magnetization plateau of TLAF magnets~\cite{chubukov_1991_quantum,susuki_2013_Ba3CoSb2O9,starykh_2015_unusual}.
Importantly, these states differ in their harmonic content: the umbrella state is described by a single-$\m{Q}$ (harmonic) modulation plus a net magnetization. Meanwhile, the Y- and up-up-down states require additional higher harmonic modulations, including components at twice the primary ordering vector of the helimagnetic wave ($2\m{Q}$). 
The formation of anharmonic, or solitonic, states is enabled by the condition $J_{\m{Q}} = J_{2\m{Q}}$ for the exchange interaction in reciprocal space, which allows spin modulations at both $\m{Q}$ and $2\m{Q}$ to coexist without additional energy cost.


In this Letter, we generalize the concept of competing harmonic and anharmonic magnetic orders, induced by frustration, 
to the BCTL by combined theoretical and experimental evidence.
Within an appropriate range of exchange parameters, the condition $J_{\m{Q}} \sim J_{2\m{Q}}$ is realized for the BCTL, giving rise to competition between harmonic and anharmonic helimagnetic orders. These correspond to the cone and solitonic states, respectively.

\textit{Solitonic state and exchange interactions}~--~We use a BCTL model with intralayer nearest ($J_1$) and next-nearest-neighbor ($J_2$), and interlayer ($J_c$) exchange interactions, as shown in Fig.~\ref{Fig1}(e). When only ordering vectors in the basal plane are considered, we may project all the ions onto a single plane.
As shown in Fig.~\ref{Fig1}(f), the Fourier-transformed exchange interaction $J_{\m{k}}$ has maxima at $\m{Q}=(2/3,\,2/3,\,0)$ for reasonable model parameters $J_1=-1.0,\,J_c=1.5,\,J_2=-0.25$. This $J_{\m{k}}$ naturally satisfies the condition $J_{\m{Q}}=J_{2\m{Q}}$, similar to the case of a TLAF. 
In a finite magnetic field, two possible distortions of helimagnetic order are conical and solitonic, as shown in Fig.~\ref{Fig1}(g,h). Both have a finite net magnetization, but the latter is realized by adding higher harmonic content to the helimagnetic wave. 
The energies of these two distorted helimagnetic states, $\mathcal{E}_\mathrm{cone}$ and $\mathcal{E}_\mathrm{sol}$, in the BCTL model are described as
\begin{equation}
\begin{aligned}
\mathcal{E}_{\mathrm{cone}} 
&\propto -\frac{S^2  J_{{\m{Q}}}  }{2}
- \frac{h^2}{2\left( J_{{\m{Q}}} - J_\m{0}  \right)},\\
\mathcal{E}_{\mathrm{sol}}
&\propto -\frac{S^2  J_{{\m{Q}}}  }{2}
- \frac{h^2}{2\left( 2J_{{\m{Q}}} - J_\m{0} - J_{{2\m{Q}}} \right)},
\label{Eq_suscept}
\end{aligned}
\end{equation}
where $J_\m{0}$ and $h$ are the exchange interaction at $\m{k}=\m{0}$ and the in-plane magnetic field in units of energy, respectively~\cite{zhitomirsky1996,Utesov_2018_suscept}. 
Exchange anisotropy and Dzyaloshinskii-Moriya (DM) interactions produce corrections to these energies~\cite{SM}, however, when $J_{\m{Q}}- J_{2\m{Q}}$ is much larger than the energy scale of such corrections, the ordering vector $\m{Q}$ always realizes a cone state.

Note that a solitonic state, termed `chiral soliton lattice', has been reported in monoaxial chiral helimagnets, where the ordering vector $\m{Q}$ is fixed along a single direction. In this chiral material platform, the spin rotation plane is fixed by the Dzyaloshinskii–Moriya interaction and a magnetic field applied within the plane of spin rotation 
results in the formation of a solitonic state. The present geometry is different: $\m{Q}$ can point along multiple directions in the basal plane, the magnetic field is also applied within the plane, and naively the conical state should be favored~\cite{kishine_2005_synthesis,togawa_2012_CSL,togawa_2013_CNS,matsumura_2017_CSL}. Nevertheless, the conical and solitonic states become degenerate, when the frustration condition $J_{\m{Q}}=J_{2\m{Q}}$ is relevant in the BCTL.

\begin{figure}[!htp]
  \centering
  \includegraphics[width=0.99\linewidth]{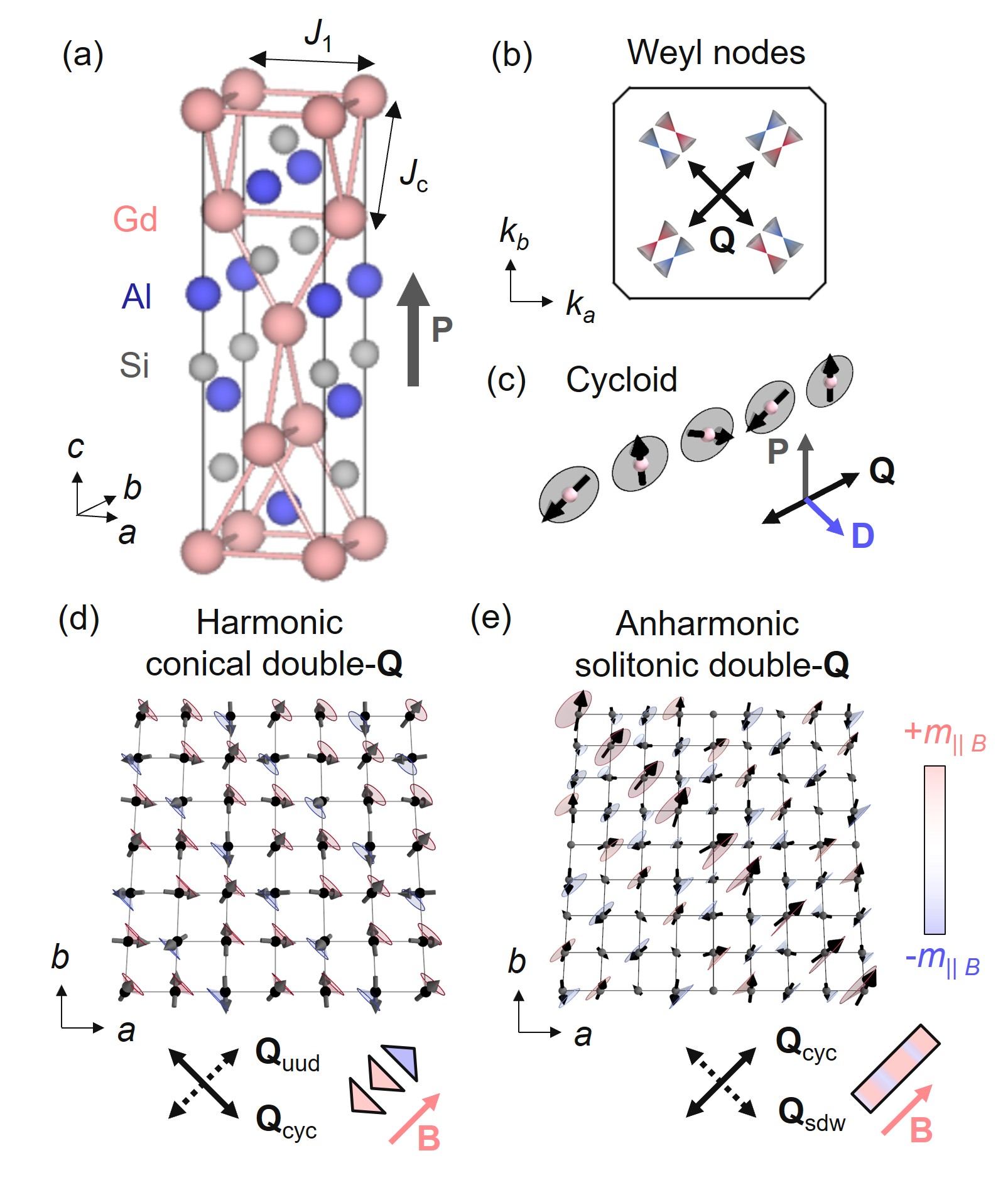} 
\caption{(color online) Competing magnetic orders in the Weyl helimagnet GdAlSi with BCTL of magnetic ions.
(a) Tetragonal crystal structure with the polar axis $\m{P}$ along the $c$-axis. Frustration arises from competing nearest intralayer ($J_1$) and interlayer ($J_c$) interactions.
(b) Brillouin zone (BZ) of GdAlSi projected onto the $k_c = 0$ plane. Weyl nodes are connected by the magnetic ordering vector $\m{Q}$.
(c) Cycloidal helimagnetic structure in the zero-field ground state of GdAlSi. This state is consistent with Dzyaloshinskii–Moriya (DM) interactions allowed by the polar crystal structure.
(d,e) Competing magnetic phases of GdAlSi under a high magnetic field $\m{B}$ applied along the $[110]$-direction. These consist of a superposition of two magnetic waves, forming a double-$\m{Q}$ state. We observe a transition between helimagnetic cycloid (d) and anharmonic solitonic cycloid (e), tunable by $\m{B}$ and temperature. The red and blue colors represent the moment parallel and antiparallel to $\m{B}$, respectively.
The conical double-$\m{Q}$ state consists of two cones parallel to $\m{B}$ and one cone antiparallel to it. Meanwhile, the solitonic double-$\m{Q}$ state involves anharmonic spin tilting towards $\m{B}$.
}
  \label{Fig2}
\end{figure}

\textit{Helimagnetic Weyl semimetal GdAlSi}~--~To demonstrate this type of frustration physics --- competing harmonic and anharmonic states --- in the BCTL, we reveal conical and solitonic helimagnetic orders in the magnetic Weyl semimetal GdAlSi, which has the suitable magnetic ordering vector $\m{Q}\approx(2/3,\,2/3,\,0)$. Figure~\ref{Fig2}(a) shows GdAlSi's noncentrosymmetric crystal structure, in which the polar axis is oriented along the $c$-axis. In GdAlSi's electronic structure, topological Weyl fermions emerge along the directions equivalent to $[110]$ in the $k_c=0$ plane, as schematically illustrated in Fig.~\ref{Fig2}(b)~\cite{laha_2024_GdAlSi,Nag_2024_GdAlSi,nakano_2026_GdAlSi}. The magnetic Gd moments begin to order at $32\,\mathrm{K}$, below which a cycloidal helimagnetic order is realized in zero magnetic field, shown in Fig.~\ref{Fig2}(c).

The zero-field magnetic cycloid in GdAlSi is understood in two steps. First, the intralayer ($J_1$) and interlayer ($J_c$) nearest-neighbor interactions compete on the Gd sublattice [Fig.~\ref{Fig2}(a)]. Like the case of Fig.~\ref{Fig1}(a), the Fourier-transform $J_{\m{k}}$ is maximized at $\m{Q}=(2/3,\,2/3,\,0)$ and yields $J_{\m{Q}}=J_{2\m{Q}}$ in the parameter regime appropriate for GdAlSi, $J_c = 2J_1$. Indeed, an almost commensurate magnetic ordering vector, $\m{Q}=(q,\,q,\,0)$ with $q=0.673$, has been observed experimentally in GdAlSi~\cite{nakano_2026_GdAlSi}.

Once the ordering vector $\m{Q}$ is selected, the DM interaction, allowed by the polar crystal structure, becomes important to determine the direction of the magnetic moments --- or the rotation plane of the helimagnetic wave. In polar GdAlSi, the DM vectors are perpendicular to both the polar axis and $\m{Q}$, favoring the spin-rotation plane to be cycloidal, as shown in Fig.~\ref{Fig2}(c).

\begin{figure}[!htp]
  \centering
  \includegraphics[width=1.05\linewidth]{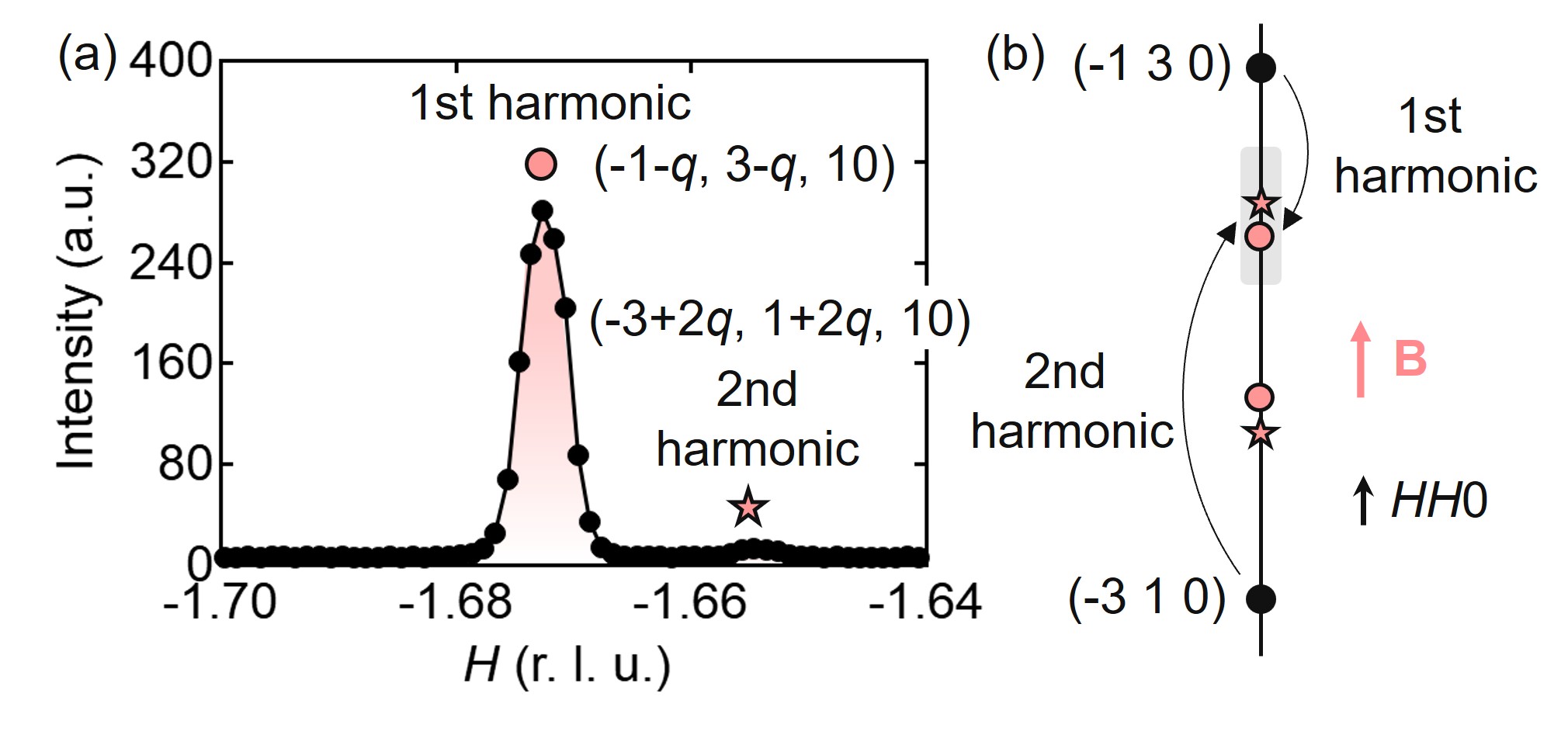} 
  \caption{(color online)
Solitonic distortion of the cycloidal helimagnetic order in GdAlSi
revealed by resonant elastic X-ray scattering (REXS) measurements.
(a) Magnetic scattering from the second-order harmonic reflection $(-3+2q,\,1+2q,\,10)$, demonstrating an anharmonic distortion consistent with a solitonic cycloidal texture. The experiments are conducted in the solitonic double-$\m{Q}$ state at $27\,\mathrm{K},\,7\,\mathrm{T}$.
(b) Schematic of the $HH0$ line in reciprocal space. The line scan shown in panel (a) is highlighted in grey, passing through the first harmonic of $(-1,\,3,\,0)$ (red circle) and the second harmonic of $(-3,\,1,\,0)$ (red star).
    }
    \label{Fig3}
\end{figure}

\textit{Competing harmonic and solitonic states in GdAlSi}~--~
A magnetic field applied along the $[110]$ direction transforms the ground state of GdAlSi into one of two competing, distorted helimagnetic states: one of conical, and one of solitonic character. Both these states are superpositions of two magnetic waves termed double-$\m{Q}$ states: an additional, collinear modulation appears normal to the main helimagnetic wave~\cite{SM}.

Figure~\ref{Fig2}(d) shows the magnetic ground state of GdAlSi at low temperatures and high magnetic fields. This is the conical double-$\m{Q}$ state~\cite{nakano_2026_GdAlSi}. The cycloidal $\m{Q}_\mathrm{cyc}$ remains the dominant ordering vector, being oriented perpendicular to $\m{B}$. However, this cycloid is distorted into staggered cones, corresponding to a secondary, commensurate modulation with an up-up-down spin configuration. Despite this secondary component, the primary cycloid remains nearly harmonic. 

In this work, we identify a solitonic double-$\m{Q}$ state emerging in a magnetic field at elevated temperatures, which has not been reported so far~\cite{nakano_2026_GdAlSi}. This state is shown in Fig.~\ref{Fig2}(e). Our first important observation is that, in the solitonic double-$\m{Q}$ phase, the dominant cycloidal-$\m{Q}_\mathrm{cyc}$ aligns parallel to the field direction. This is striking and unconventional, because $\m{Q}_\mathrm{cyc}\parallel \m{B}$ does not allow for a conical distortion. Instead, the spin rotation becomes anharmonic, resulting in a finite net magnetization within the cycloid plane.

\textit{X-ray diffraction experiments}~--~Experimentally, the anharmonicity of the solitonic texture manifests itself as second-harmonic reflections in magnetic diffraction measurements. We perform resonant elastic X-ray scattering (REXS) experiments at the Gd $L_2$ absorption edge, details of which are given in the Supplemental Material~\cite{SM}. Figure~\ref{Fig3}(a,b) shows a REXS line scan parallel to the direction of the applied magnetic field. The main peak visible here is the magnetic reflection $(-1-q,\,3-q,\,10)$, which is a magnetic satellite of the Bragg reflection $(-1,\,3,\,10)$. Right next to it, a weaker magnetic peak is observed at $(-3+2q,\,1+2q,\,10)$, and this is identified as a second-harmonic satellite of the $(-3,\,1,\,10)$ Bragg reflection. The observation of this second-harmonic reflection is the main experimental result of our paper, and an incisive proof for the field-induced solitonic texture in GdAlSi. In the End Matter, we provide more details of the experiments, including polarization analysis of the scattered REXS intensity.

\begin{figure}[!htp]
  \centering
  \includegraphics[width=1.0\linewidth]{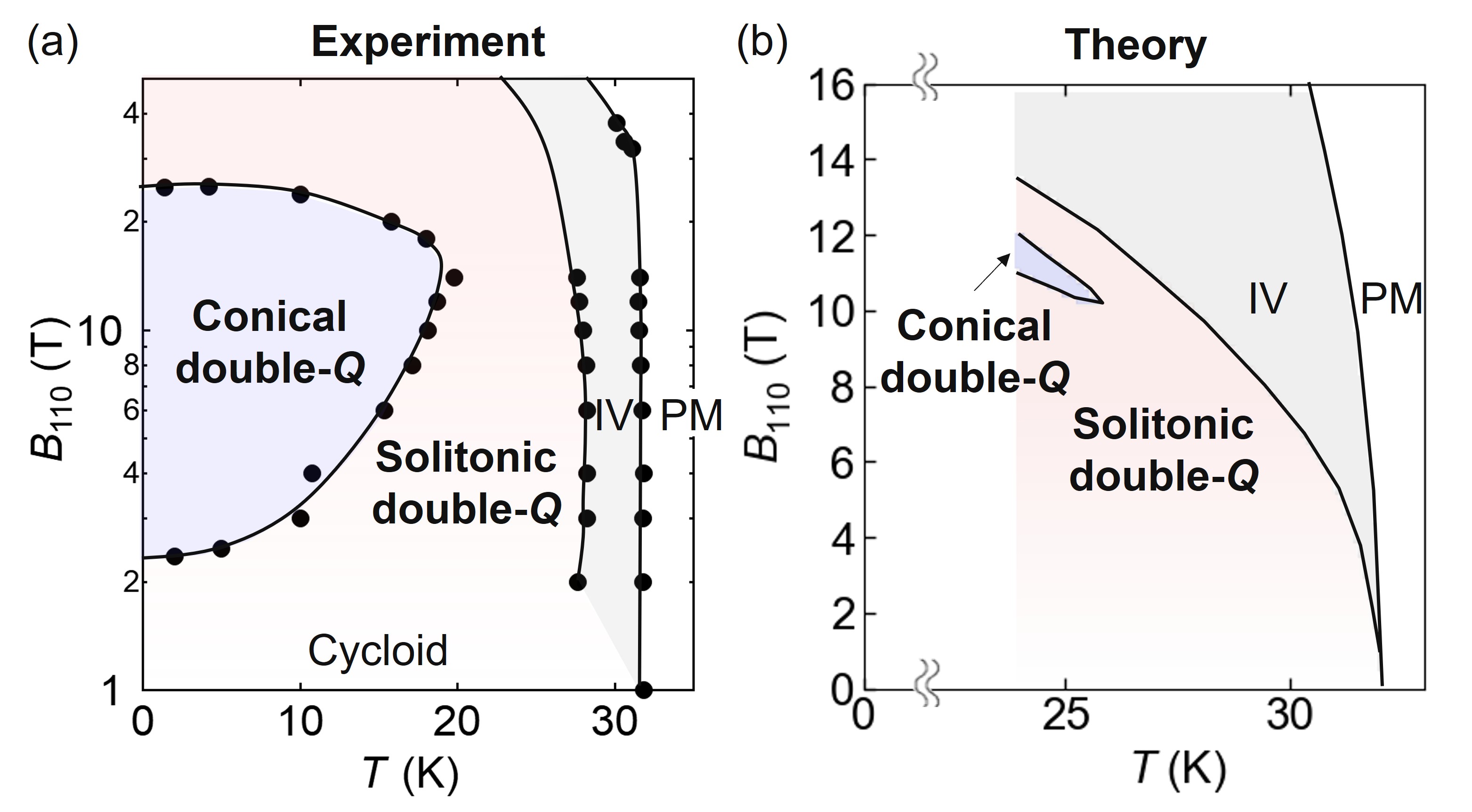} 
  \caption{(color online) 
(a) Experimental magnetic phase diagram of GdAlSi with $\m{B}\parallel[110]$, determined from magnetization experiments~\cite{SM}. The conical and solitonic double-$\m{Q}$ states compete under magnetic fields, with the conical double-$\m{Q}$ state exhibiting a pocket-like region.
(b) Magnetic phase diagram from mean-field theory, considering frustrated magnetic interactions with $J_{\m{Q}}\sim J_{2\m{Q}}$ on the BCTL of GdAlSi. The theory well reproduces the essential features of the experimental phase diagram at high temperatures.}
    \label{Fig4}
\end{figure}

\textit{Modeling the phase diagram}~--~We consider the magnetic phase diagram of GdAlSi with $\m{B}\parallel[110]$.
This phase diagram prominently features the competing double-$\m{Q}$ states discussed above: a conical double-$\m{Q}$ and a solitonic double-$\m{Q}$ state. The phase boundaries in Fig.~\ref{Fig4}(a) are determined by magnetization measurements in pulsed magnetic fields up to $60\,\mathrm{T}$~\cite{SM}. The harmonic and anharmonic double-$\m{Q}$ states compete at finite fields, and there is a pocket-shaped region for the conical double-$\m{Q}$ state at low temperature. 

The magnetic phase diagram of GdAlSi is well reproduced by calculations in mean-field theory, shown in Fig.~\ref{Fig4}(b). The key ingredients of the model are tetragonal symmetry --- 
allowing for two symmetry-equivalent ordering vectors in the basal plane --- and  $J_{\m{Q}} \approx J_{2\m{Q}}$. Both of these ingredients are a natural consequence of $\m{Q}\approx (2/3, 2/3, 0)$ on the BCTL. The details of the resulting phase diagram, with its subtle balance of harmonic and solitonic phases in a magnetic field, are also affected by  anisotropic interactions: the Dzyaloshinskii-Moriya interaction (DMI) and the anisotropic exchange interaction~\cite{SM}. 
At the highest temperature, a single-$\m{Q}$ harmonic conical state is favored by DMI (phase IV). As compared to Ref.~\cite{nakano_2026_GdAlSi}, the present model incorporates the effect of $J_{2\m{Q}}$ and consequently realizes solitonic distortions at finite temperature. We confirmed that the solitonic state is absent in the model of Ref.~\cite{nakano_2026_GdAlSi}. 

\textit{Conclusions}~--~By a combination of theory and experiments we demonstrate that concepts from frustration physics on the triangular lattice can be applied to the BCTL, which is little explored to date~\cite{mathur_1998_BCTL, sebastian_2006_BCTL,farias_2016_BCTL}. The presence of degenerate ordering vectors, specifically $\m{Q}$ and $2\m{Q}$, provides a fresh perspective for complex magnetic orders on the square lattice~\cite{khanh_2022_zoology}. In the case of GdAlSi, this frustration condition plays an essential role in stabilizing the anharmonic magnetic state. The relationship between harmonic and anharmonic helimagnetic structures realized in GdAlSi, and Weyl electrons in its band structure, remains to be clarified. In particular, future studies should address whether and how much Weyl electrons contribute to the Dzyaloshinskii-Moriya and anisotropic exchange interactions in this class of topological semimetal.

\vskip\baselineskip
\textbf{Acknowledgements}\\ 
We acknowledge fruitful discussions with Shota Suetsugu and Moyu Kato.
R.N. was supported by the Program for Leading Graduate Schools (MERIT-WINGS), and JST SPRING, Grant Number JPMJSP2108.
O.I.U. and S.K.K. were supported by the Brain Pool Plus Program through the National Research Foundation of Korea, funded by the Ministry of Science and ICT (2020H1D3A2A03099291). O.I.U. and S.K.K. were also funded by the National Research Foundation of Korea (NRF), funded by the government of Korea (MSIT), Grant No. RS-2026-25470048; they are also funded by the Creation of the quantum information science R\&D ecosystem (based on human resources) through the National Research Foundation of Korea (NRF), funded by the Korean government (Ministry of Science and ICT(MSIT)), Grant No. RS-2023-00256050.
M.H. is supported by the Deutsche Forschungsgemeinschaft (DFG, German Research Foundation) via Transregio
TRR 360 - 492547816. 
This work was supported by JSPS KAKENHI Grants No. 25K17336, No. 26H00587, No. 26H01294, No. JP26H00644, and No. JP26H01290. This work was partially supported by the Japan Science and Technology Agency via JST CREST Grant No. JPMJCR20T1 (Japan), JST PRESTO JPMJPR259A, and JST FOREST JPMJFR2238. It was also supported by Japan Science and Technology Agency (JST) as part of Adopting Sustainable Partnerships for Innovative Research Ecosystem (ASPIRE), Grant No. JPMJAP2426.\\

\bibliography{reference}

\begin{widetext}
\end{widetext}

\begin{center}
\textbf{End Matter}
\end{center}

\begin{figure}[H]
  \centering
  \includegraphics[width=1.00\linewidth]{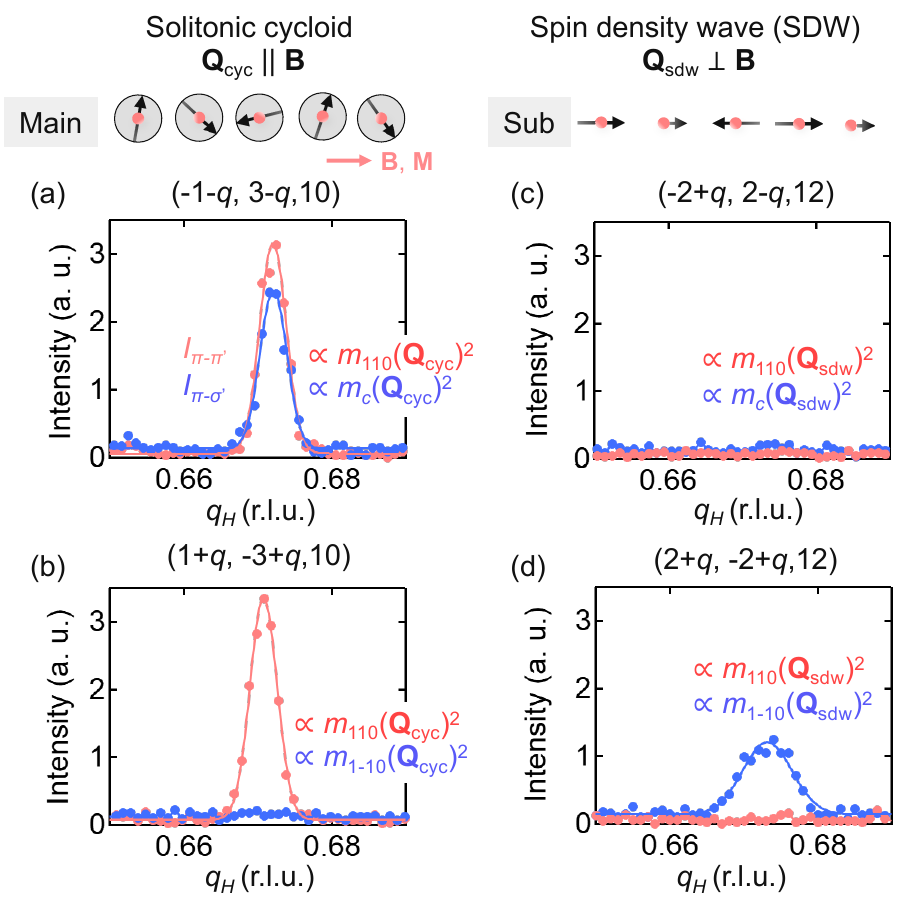} 
  \caption{
Polarization analysis of REXS in the solitonic double-$\m{Q}$ state at $27\,\mathrm{K},\,7\,\mathrm{T}$.
(a.b) Line profiles of $(-1-q_\mathrm{H},\, 3-q_\mathrm{H},\, 10)$ and $(1+q_\mathrm{H},\, -3+q_\mathrm{H},\, 10)$ scans for $0.65 < q_\mathrm{H} < 0.69$ around the solitonic cycloidal reflection ($\m{Q}_{\mathrm{cyc}} \parallel \m{B}$).
Two inequivalent reflection positions, $(-1-q,\, 3-q,\, 10)$ and $(1+q,\, -3+q,\, 10)$, are selected such that the intensity $I_{\pi\text{-}\sigma'}$ from the $\pi$--$\sigma'$ channel predominantly probes $m_{c}(\m{Q}_{\mathrm{cyc}})^{2}$ and $m_{1\overline{1}0}(\m{Q}_{\mathrm{cyc}})^{2}$, respectively, whereas $I_{\pi\text{-}\pi'}$ from the $\pi$--$\pi'$ channel probes $m_{110}(\m{Q}_{\mathrm{cyc}})^{2}$ for both reflections~\cite{SM}. Sharp peaks observed in the $\pi$--$\pi'$ channel for both reflections, together with a peak in the $\pi$--$\sigma'$ channel only at $(-1-q,\, 3-q,\, 10)$, indicate the presence of finite $m_{110}(\m{Q}_{\mathrm{cyc}})$ and $m_{c}(\m{Q}_{\mathrm{cyc}})$ components, thereby demonstrating the cycloidal spin character.
(c,d) Line profiles of $(-2+q_\mathrm{H},\, 2-q_\mathrm{H},\, 12)$ and $(2+q_\mathrm{H},\, -2+q_\mathrm{H},\, 12)$ scans for $0.65 < q_\mathrm{H} < 0.69$ around the collinear spin-density wave (SDW) reflections ($\m{Q}_{\mathrm{sdw}} \parallel \m{B}$). 
Sharp peaks in the $\pi$--$\sigma'$ channel for $(2+q,\,-2+q,\,12)$ reflections indicate a finite $m_{1\text{-}10}(\m{Q}_{\mathrm{sdw}})$ components.
}
    \label{Fig5}
\end{figure}

\textit{Polarization analysis in resonant X-ray diffraction for the solitonic double-Q state}~--~To provide a detailed magnetic structure model in the solitonic double-$\m{Q}$ state, we perform polarization analysis of the resonantly scattered X-rays in Fig.~\ref{Fig5}(a-d). In our experiments, the incident X-rays are linearly polarized parallel to the scattering plane ($\pi$-polarization). The scattered X-ray beam contains both components parallel ($\pi'$) and perpendicular ($\sigma'$) to the scattering plane, which are separated using an analyzer plate~\cite{SM}. In our experimental geometry, the intensity $I_{\pi\text{-}\pi'}$ corresponds to the Fourier component of magnetic moments along the $[110]$-direction, whereas $I_{\pi\text{-}\sigma'}$ probes the component within the $(H\text{-}HL)$ plane. 
For ordering vectors along the two directions equivalent to $[110]$ in the tetragonal structure of GdAlSi, we select two representative reflections such that the Fourier components along $c$ and $[1\text{-}10]$ are mapped onto the $I_{\pi\text{-}\sigma'}$ channel of each reflection~\cite{SM}.
Figure~\ref{Fig5}(a,b) shows the polarization analysis of the dominant ordering vector $\m{Q}$, which is along the direction of the external magnetic field. Sharp peaks in the $\pi$--$\pi'$ channel for both reflections indicate a finite $[110]$-component of the ordered magnetic moment, while a pronounced peak in $I_{\pi\text{-}\sigma'}$ --- observed only for $(-1-q,\,3-q,\,10)$ --- demonstrates the presence of a $c$-component. In combination, these results are characteristic of a cycloidal magnetic structure for this $\m{Q}$. 

Along the direction orthogonal to the external magnetic field, we observe an incommensurate modulation at $\m{Q}_\mathrm{sdw}$ with, relatively speaking, shorter period. Polarization analysis in Fig.~\ref{Fig5}(c,d) indicates that only the $[1\text{-}10]$-component is present. This is consistent with a collinear spin-density wave (SDW) structure, shown in the right-top of Fig.~\ref{Fig5}.

Finally, the solitonic double-$\m{Q}$ state illustrated in Fig.~\ref{Fig2}(e) is realized as a superposition of two $\m{Q}$-vectors with, respectively, cycloidal and SDW spin modulations.

\end{document}